# Stacked Intelligent Metasurfaces for Resolution-Constrained Near-Field Range Extension in 6G Systems

Yajun Zhao, *Member, IEEE*

***Abstract*—Near-field electromagnetic focusing is central to 6G communication, sensing, and integrated sensing and communication (ISAC) systems. However, for a fixed aperture, the resolution-constrained usable range of conventional single-layer transmissive metasurfaces is far shorter than the classical Rayleigh distance. This discrepancy stems not from fundamental near-field physics limitations, but from inadequate wavefront control, implementation imperfections, and the quadratic degradation of axial resolution with distance.To quantify this gap, we distinguish the Rayleigh distance from the engineering-usable near-field distance (UNFD), defined as the maximum range where predefined focusing gain and resolution requirements are jointly satisfied. Under identical aperture, feed, and input power constraints, we investigate how stacked intelligent metasurfaces (SIMs) extend UNFD via cascaded wavefront shaping. A unified framework combining effective-phase-distance and discrete Green's-function operator perspectives is developed, interpreting multilayer SIM design as an operator approximation problem for ideal near-field focusing.We derive low-complexity analytical models revealing an inherent distance-resolution dilemma: lateral resolution degrades linearly with distance, while axial resolution degrades quadratically, making axial performance the dominant bottleneck. Multilayer stacking mitigates this by improving wavefront curvature matching and reducing residual phase error. Engineering correction factors for practical imperfections and a higher-order phase framework for extreme near-field operation are incorporated. Numerical simulations confirm that increasing layer count pushes UNFD closer to the Rayleigh limit, but gains saturate as accumulated losses offset control flexibility benefits.**

***Index Terms*—Stacked intelligent metasurface, engineering-usable near-field distance, spatial resolution, multilayer wavefront shaping.**

## I. INTRODUCTION

THE evolution of 6G wireless systems toward higher carrier frequencies, electrically larger apertures, and deeply integrated communication and sensing functionalities has rendered numerous practical application scenarios fall within the radiative near-field regime [1][2]. In this regime, the conventional far-field plane-wave approximation fails to capture the essential propagation characteristics. Instead, electromagnetic fields exhibit prominent spherical-wave features, and system responses depend jointly on both azimuth angle and propagation distance. This unique property enables distance-selective beam focusing, high-precision localization, and distance-angle coupled spatial multiplexing, all of which are regarded as core capabilities for future 6G networks [3].

Reconfigurable intelligent surfaces (RISs) and transmissive programmable metasurfaces have emerged as promising hardware platforms for realizing the above functionalities, owing to their ability to reshape electromagnetic wavefronts with low power consumption and hardware complexity [4]. However, most existing designs rely on a single programmable layer. Although single-layer transmissive metasurfaces can emulate programmable thin lenses and achieve basic near-field focusing, their field transformation capabilities are fundamentally limited by the single-plane modulation architecture. This limitation becomes particularly prominent in cases of long focal distances, weak required wavefront curvature, nonuniform feed illumination, and non-negligible implementation imperfections [6].

Stacked intelligent metasurfaces (SIMs), consisting of multiple cascaded transmissive programmable layers, offer a more flexible field manipulation mechanism [5][7]. Through iterative free-space propagation and electromagnetic modulation, SIMs expand the set of practically realizable wavefront transformations under a fixed physical aperture, making them highly suitable for near-field focusing tasks requiring accurate synthesis of target spherical wavefronts [8]. Nevertheless, the core practical question lies not in whether multilayer stacking increases the peak field intensity at the target point, but in whether it can effectively extend the usable near-field operating distance under stringent focusing, resolution, and robustness constraints.

In classical antenna theory, the radiative near-field boundary is conventionally defined by the Rayleigh distance $R_{\mathrm{Ray}} = 2D^2/\lambda$, where $D$ denotes the aperture diameter and $\lambda$ denotes the operating wavelength. As a geometric-physical upper bound for the classical near-field region under fixed-aperture assumptions, the Rayleigh distance depends solely on aperture size and wavelength [11]. In practical implementations, however, metasurface systems often fail to maintain satisfactory focusing quality across the entire range $(0, R_{\mathrm{Ray}}]$. In other words, the practical usable operating range is considerably shorter than the theoretical physical near-field boundary.

Yajun Zhao is with ZTE Corporation, Beijing 100029, P.R.China, and State Key Laboratory of Mobile Network and Mobile Multimedia Technology , Shenzhen 518055, P.R.China. (e-mail: zhao.yajun1@zte.com.cn ).

To clarify this critical distinction, this paper introduces the concept of engineering-usable near-field distance (UNFD). Unlike the Rayleigh distance — a physical boundary determined by inherent aperture characteristics — the UNFD is an application-oriented engineering metric. It is defined as the maximum propagation distance within the classical near-field region at which a metasurface system can simultaneously meet predefined requirements for focusing gain, lateral resolution, and axial resolution. Under this definition, the core function of SIMs is not to alter the fundamental physical near-field boundary, but to extend the effective operating range to approach the Rayleigh distance as closely as possible.

The necessity of this distinction stems from the inherent spatial resolution characteristics of near-field systems. For a fixed aperture size, the ideal diffraction-limited lateral resolution is approximately proportional to $\lambda r/D$, while the axial resolution scales with $\lambda r^2/D^2$. As the focal distance $r$ increases, axial resolution degrades far more rapidly than lateral resolution, leading to an inherent distance-resolution dilemma: even when the system operates physically within the classical near-field region, its distance-domain selectivity may already fail to meet engineering requirements. From the perspective of electromagnetic information theory (EIT), this indicates that the practically usable mode density in the distance domain drops sharply with increasing propagation distance. Consequently, the three-dimensional performance advantages of near-field propagation diminish long before reaching the Rayleigh boundary.

Existing studies on SIMs have verified that additional layers can enhance beamforming flexibility and peak focusing gain at target points [9][10]. However, a systematic framework that quantitatively correlates the number of stacked layers, wavefront approximation accuracy, spatial resolution retention capability, and engineering-usable near-field distance remains missing. Moreover, most existing analyses rely solely on phase profile characterization, lacking a unified field propagation framework to interpret gain degradation, residual phase errors, point-spread function broadening, and operator mismatch in a consistent manner.

Driven by the above research gaps, this paper investigates the mechanism by which SIMs extend the UNFD under fixed-aperture constraints. The core principle is that multilayer stacking improves the approximation accuracy of target spherical wavefronts via cascaded modulation and free-space propagation, thereby reducing residual phase errors and delaying the degradation of lateral and axial resolutions. To formalize this mechanism, we develop a unified framework integrating the effective phase-distance model and discrete Green's function operator model. The former provides intuitive engineering metrics such as residual curvature mismatch, while the latter reformulates multilayer near-field focusing as an operator approximation problem targeting the ideal near-field focusing operator.

In our previous work [14], we investigated near-field effective range extension using SIMs from the perspective of beamforming focusing gain loss. In contrast, this paper specifically addresses resolution-constrained near-field range extension via stacked intelligent metasurfaces, which represents the core novelty of the present study. This resolution-constrained engineering-usable near-field distance (UNFD) is a critical performance metric for 6G integrated sensing and communication (ISAC) systems [15][16].

The main contributions of this paper are summarized as follows.

1) We explicitly distinguish the classical Rayleigh distance from the proposed UNFD, and establish a threshold-based practical UNFD definition considering joint constraints of focusing gain and spatial resolution.

2) We reveal the inherent distance-resolution dilemma in near-field focusing systems, and validate that axial resolution acts as the dominant resolution constraint limiting long-range near-field operation under fixed aperture conditions.

3) We develop a unified multilayer modeling framework based on effective phase distance and discrete Green's function operators, and reinterpret SIM design as an approximation problem for ideal near-field focusing operators.

4) We provide an in-depth physical interpretation of the multilayer wavefront approximation mechanism, demonstrating that the performance gains of stacking originate from alternating local modulation and nonlocal propagation, rather than a simple increase in phase control variables.

The remainder of this paper is organized as follows. Section II presents the system model and problem formulation. Section III elaborates on the wavefront characterization, spatial resolution analysis, and unified operator-based framework, including detailed derivations from Fresnel propagation to high-order phase approximation. Section IV provides numerical simulations and performance evaluations. Section V concludes the paper.

## II. SYSTEM MODEL AND PROBLEM FORMULATION

### *A. SIM Configuration and Fair-Comparison Principle*

We consider a transmissive SIM consisting of $L$ parallel programmable metasurface layers spaced by a distance $d$. Each layer comprises $N$ subwavelength transmissive elements arranged within an identical physical aperture. A feed source placed behind the first layer illuminates the layered stack, and the cascaded structure is designed to focus the transmitted electromagnetic field at a target point $P = (0,0,r)$ on the system boresight.

To ensure a fair comparison among configurations with different layer numbers, all setups adopt identical physical aperture, feed geometry, element spacing, operating frequency, and total input-power normalization. Accordingly, all performance discrepancies arise from the enhanced field-transformation flexibility provided by longitudinal stacking, rather than from enlarged aperture size or additional transmit power. Unless otherwise stated, $D$ denotes the diameter of an equivalent circular aperture having the same active physical area as the metasurface region, which allows standard Rayleigh-distance formulas to be consistently used throughout the paper.

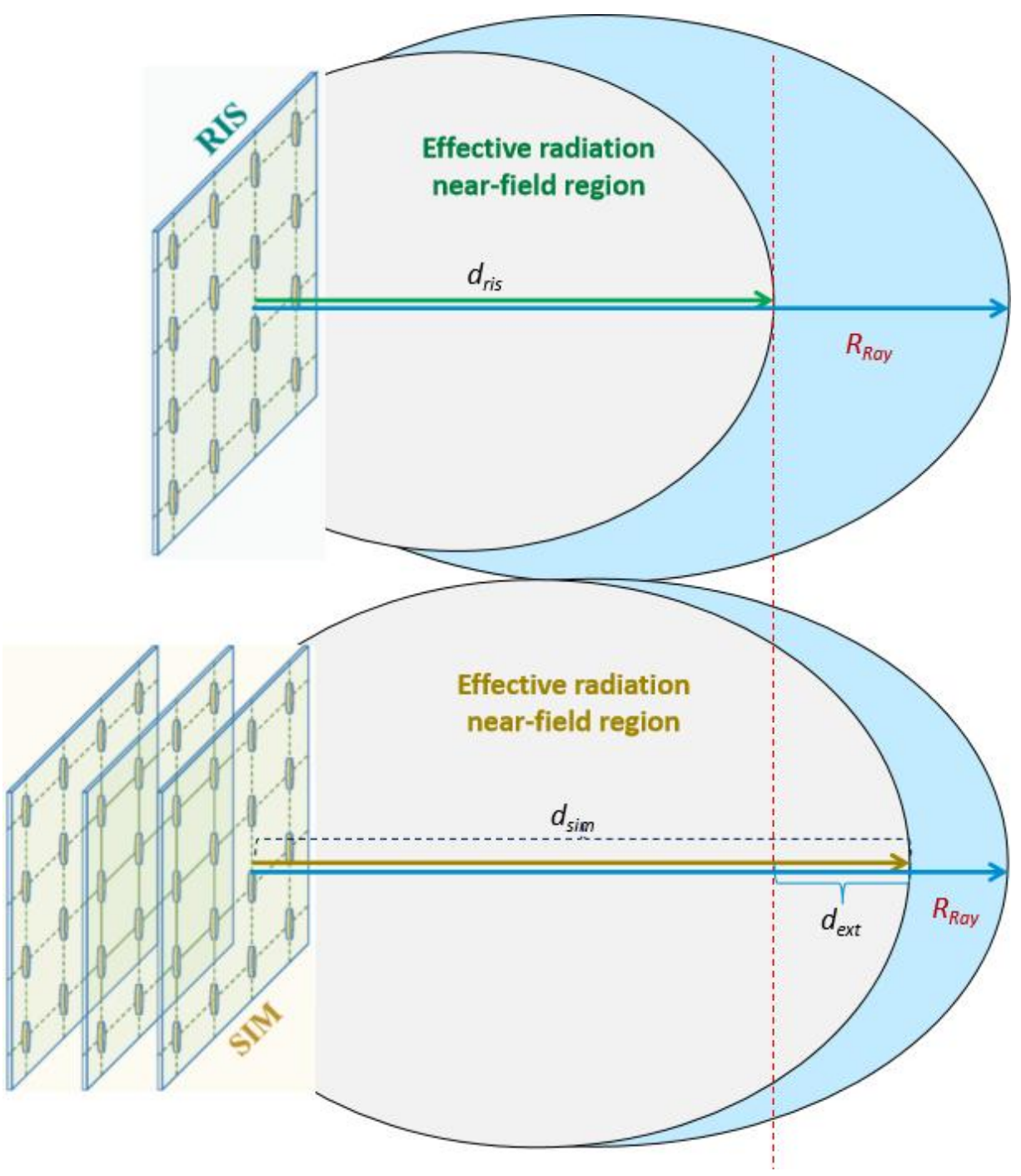

**Fig. 1.** Schematic illustration of the radiative near field for a single-layer metasurface and a transmissive stacked intelligent metasurface. (Schematic illustration of the radiative near-field regions for a single-layer metasurface (UNFD = $d_{RIS}$ ) and a transmissive stacked intelligent metasurface (UNFD = $d_{SIM}$ ), where $d_{ext} = d_{SIM} - d_{RIS}$ denotes the extended usable range achieved by SIMs.)

Let the complex transmission coefficient of the $n$-th element on the l-th layer be expressed as

$$\Gamma_n^{(l)} = A_n^{(l)} e^{j\phi_n^{(l)}}, \quad (1)$$

where $A_n^{(l)}$ and $\phi_n^{(l)}$ represent the amplitude response and phase response of the element, respectively. For an ideal lossless phase-only modulation scenario, $A_n^{(l)} = 1$ . In practical engineering implementations, however, $A_n^{(l)}$ may deviate from unity due to insertion loss and fabrication tolerances.

### *B. Discrete Green's-Function Representation*

Under the scalar-wave approximation, field propagation between adjacent layers can be described by the free-space Green's function [17]: $G(r, r') = \frac{e^{-jk_0|r-r'|}}{4\pi|r-r'|}$, where $k_0 = 2\pi/\lambda$ is the free-space wavenumber. After discretization over individual metasurface elements, the propagation process from layer $l$ to layer $l+1$ can be characterized by a propagation matrix, $G_l \in \mathbb{C}^{N\times N}$ . Meanwhile, the programmable electromagnetic response of the $l$-th layer is formulated as a diagonal matrix, $\Gamma_l = \mathrm{diag}(\Gamma_1^{(l)}, \Gamma_2^{(l)}, ..., \Gamma_N^{(l)})$.

Accordingly, the effective field transformation from the incident field on the first layer to the output field immediately after the last layer is expressed as:

$$T_{SIM}^{(L)} = G_{L-1}\Gamma_{L-1} \cdots G_2\Gamma_2 G_1\Gamma_1. \quad (2)$$

Let $H_\Omega(r) \in \mathbb{C}^{Q\times N}$ denote the propagation matrix mapping the field from the aperture of the last layer to a sampled observation region $\Omega_r$ centered at the target focal distance $r$. The overall effective focusing operator of the stacked SIM system is then defined as:

$$G_{SIM}^{(L)}(r) = H_\Omega(r) T_{SIM}^{(L)}. \quad (3)$$

This operator representation is crucial because it explicitly distinguishes two fundamentally different ingredients: the metasurface layers provide programmable diagonal modulation, whereas free-space propagation provides dense nonlocal field mixing. Therefore, a multilayer SIM is not merely a set of additional phase masks, but a cascaded modulation-propagation architecture whose end-to-end approximation capability is substantially richer than that of a single-layer structure.

### *C. Effective Phase Distance and Residual Phase Error*

In addition to the operator-based perspective, a phase-domain engineering description is introduced to facilitate physical interpretation. We consider a complete propagation path from the feed source to the focal target point. To eliminate ambiguities in phase sign conventions, we define the total accumulated phase along an $L$-layer propagation path as $\Phi_{tot,L}(r) = k_0 d_{prop} + \sum_{l=1}^{L} \phi_l^{(comp)} + \phi_{Rx}^{(comp)}$, where $d_{prop}$ is the geometric free-space propagation path length, $\phi_l^{(comp)}$ represents the compensation phase provided by the $l$-th metasurface layer, and $\phi_{Rx}^{(comp)}$ denotes the additional compensation phase applied at the observation terminal.

Correspondingly, the effective propagation distance is formally defined as

$$r_{eff,L}(r) \triangleq \frac{\Phi_{tot,L}(r)}{k_0} = d_{prop} + \frac{1}{k_0}\sum_{l=1}^{L} \phi_l^{(comp)} + \frac{1}{k_0}\phi_{Rx}^{(comp)}. \quad (4)$$

Ideal near-field focusing is achieved when the effective propagation distances of all aperture sampling paths are equalized at the target point, such that all residual phase deviations vanish modulo $2\pi$.

The residual phase error is therefore formulated as

$$\Delta\phi_{res}^{(L)} = \Phi_{tot,L}^{(actual)} - \Phi_{tot}^{(ideal)} = k_0(r_{eff,L}^{(actual)} - r_{eff}^{(ideal)}). \quad (5)$$

The residual phase error is taken modulo $2\pi$ to account for the periodic nature of electromagnetic phase, i.e., $\Delta\phi_{res}^{(L)} = (\Phi_{tot,L}^{(actual)} - \Phi_{tot}^{(ideal)}) \mathrm{mod} 2\pi$.

This metric provides a direct link between wavefront mismatch and practical performance degradation, including focal-gain loss, point-spread-function broadening, and resolution deterioration.

### *D. Ideal Target Near-Field Response*

For an on-axis target point $P = (0,0,r)$, let $\rho_n$ denote the radial coordinate of the $n$-th aperture sampling element. The ideal near-field focusing characteristic is described by the target field vector $g_{target}(r) \in \mathbb{C}^{N\times 1}$, whose phase term is expressed as

$$\arg\left(g_{target,n}(r)\right) = -k_0\left(\sqrt{\rho_n^2 + r^2} - r\right). \quad (6)$$

Under the paraxial approximation where $\rho_n \ll r$, the spherical phase term can be expanded via Taylor series as

$$\sqrt{\rho_n^2 + r^2} = r + \frac{\rho_n^2}{2r} - \frac{\rho_n^4}{8r^3} + \frac{\rho_n^6}{16r^5} - \cdots, \quad (7)$$

which leads to

$$\arg\left(g_{target,n}(r)\right) \approx -\frac{k_0\rho_n^2}{2r} + \frac{k_0\rho_n^4}{8r^3} - \frac{k_0\rho_n^6}{16r^5} + \cdots. \quad (8)$$

This expansion shows that near-field focusing fundamentally requires more than a purely quadratic phase profile. The second-order term determines the dominant wavefront curvature, whereas the fourth- and higher-order terms quantify the deviation between the exact spherical wavefront and a simple parabolic approximation. These high-order terms become increasingly important when the target distance is not large compared with the aperture size.

The overall target focusing operator over the sampled focal region is denoted as $G_{target}(r)$. In this work, the multilayer SIM optimization is interpreted as a process of approximating the ideal target operator $G_{target}(r)$ as accurately as possible under practical hardware and implementation constraints.

*E. Classical Rayleigh Distance and Engineering-Usable Near-Field Distance*

For a circular aperture with diameter $D$, the maximum geometric path difference between the aperture center and edge at distance r is approximately

$$\Delta L_{max}(r) \approx \frac{D^2}{8r}. \quad (9)$$

The corresponding maximum phase difference is given by

$$\Delta\Phi_{max}(r) \approx \frac{2\pi}{\lambda}\Delta L_{max}(r) = \frac{\pi D^2}{4\lambda r}. \quad (10)$$

Setting the maximum allowable phase difference as $\Delta\Phi_{max}(r) = \pi/8$ yields the classical Rayleigh distance:

$$R_{Ray} = \frac{2D^2}{\lambda}. \quad (11)$$

Drawing on the definition philosophy of the "effective Rayleigh distance" in near-field wideband beamforming [12], We now define the engineering-usable near-field distance of an $L$ -layer SIM, denoted as $R_{UNFD}^{(L)}$, as the maximum propagation distance within the range $(0, R_{Ray}]$ that simultaneously satisfies a set of prescribed performance criteria. Three complementary evaluation metrics are adopted to constrain the usable near-field range.

1) Focusing-Gain Criterion

Let $g_{SIM}^{(L)}(r)$ represent the actual focusing response of the multilayer SIM system. The normalized field coherence between the actual response and the ideal target response is defined as

$$\eta_{coh}^{(L)}(r) = \frac{|\langle g_{SIM}^{(L)}(r), g_{target}(r)\rangle|}{\|g_{SIM}^{(L)}(r)\|_2\|g_{target}(r)\|_2}. \quad (12)$$

The focusing gain loss is further quantified as

$$L_{gain}^{(L)}(r) = -20\log_{10}\eta_{coh}^{(L)}(r). \quad (13)$$

Given a predefined allowable gain-loss threshold $L_{th}$, the gain-constrained usable distance boundary is expressed as

$$R_{gain}^{(L)} = \max\{r \le R_{Ray}: L_{gain}^{(L)}(r) \le L_{th}\}. \quad (14)$$

2) Resolution Criterion

Let $\delta_{\perp}^{(L)}(r)$ and $\delta_{\parallel}^{(L)}(r)$ denote the lateral and axial full-width-at-half-maximum (FWHM) resolutions of the SIM system, respectively. Meanwhile, $\delta_{\perp,0}(r)$ and $\delta_{\parallel,0}(r)$ represent the ideal diffraction-limited lateral and axial resolution benchmarks. The resolution retention factors are defined as

$$\Pi_{\perp}^{(L)}(r) = \frac{\delta_{\perp,0}(r)}{\delta_{\perp}^{(L)}(r)}, \Pi_{\parallel}^{(L)}(r) = \frac{\delta_{\parallel,0}(r)}{\delta_{\parallel}^{(L)}(r)}. \quad (15)$$

Given the minimum acceptable retention thresholds $\Pi_{\perp,th}$ and $\Pi_{\parallel,th}$, the resolution-constrained usable boundary is defined as

$$R_{res}^{(L)} = \max\{r \le R_{Ray}: \Pi_{\perp}^{(L)}(r) \ge \Pi_{\perp,th}, \Pi_{\parallel}^{(L)}(r) \ge \Pi_{\parallel,th}\}. \quad (16)$$

3) Residual-Phase Criterion

Under the second-order dominant approximation, the residual phase induced by wavefront curvature mismatch can be formulated as $\Delta\phi_{res}^{(L)}(\rho; r) \approx \frac{k_0}{2}\left(\frac{1}{R_{eq}^{(L)}(r)} - \frac{1}{r}\right)\rho^2$, where $R_{eq}^{(L)}(r)$ denotes the equivalent wavefront curvature radius synthesized by the multilayer SIM structure. We further define the curvature mismatch term as $\Delta C_{res}^{(L)}(r) = \frac{1}{R_{eq}^{(L)}(r)} - \frac{1}{r}$.

Accordingly, the maximum residual phase error occurring at the aperture edge is approximated by

$$\Delta\Phi_{res,max}^{(L)}(r) \approx \frac{\pi D^2}{4\lambda}|\Delta C_{res}^{(L)}(r)|. \quad (17)$$

With a specified admissible phase error threshold $\Delta\Phi_{th}$, the phase-constrained usable boundary is given by

$$R_{\phi}^{(L)} = \max\{r \le R_{Ray}: \Delta\Phi_{res,max}^{(L)}(r) \le \Delta\Phi_{th}\}. \quad (18)$$

By integrating the three aforementioned constraints, the comprehensive engineering-usable near-field distance is finally defined as

$$R_{UNFD}^{(L)} = \min\{R_{gain}^{(L)}, R_{res}^{(L)}, R_{\phi}^{(L)}\}. \quad (19)$$

This definition emphasizes that the UNFD is not a universal physical constant, but a threshold-dependent engineering metric. Its value depends on application-driven performance requirements, while its ultimate physical upper bound remains limited by the classical Rayleigh distance.

*F. Higher-Order Phase Correction in the Extreme Near Field*

When the focusing distance is extremely small compared with the aperture size, the conventional quadratic-phase approximation loses its accuracy. To quantify the valid range of the quadratic phase model, we define the truncation error between the exact spherical phase and its second-order quadratic approximation. The dominant residual term neglected in the quadratic expansion is the quartic phase component, yielding the truncation error as

$$\Delta\Phi_{trunc}^{(2)}(r) \approx \frac{k_0(D/2)^4}{8r^3} = \frac{\pi D^4}{64\lambda r^3}. \quad (20)$$

By setting the maximum tolerable truncation error as $\Delta\Phi_{trunc}^{(2)}(r) = \phi$, the lower valid boundary of the quadratic-phase model is derived as

$$R_{low}^{(2)}(\phi) = (\frac{\pi D^4}{64\lambda\phi})^{1/3}. \quad (21)$$

For the extreme near-field region satisfying $r < R_{low}^{(2)}(\phi)$, quartic phase correction is essential for accurate field modeling. Accordingly, the actual phase profile of the multilayer SIM system can be expanded as a high-order polynomial function of the radial coordinate:

$$\phi_{actual}^{(L)}(\rho; r) \approx a_2^{(L)}(r)\rho^2 + a_4^{(L)}(r)\rho^4 + \cdots, \quad (22)$$

where the ideal second-order and fourth-order coefficients corresponding to pure spherical focusing are given by

$$a_{2,ideal}(r) = -\frac{k_0}{2r}, a_{4,ideal}(r) = \frac{k_0}{8r^3}. \quad (23)$$

The quartic coefficient mismatch that characterizes the high-order phase distortion is defined as

$$\Delta a_4^{(L)}(r) = |a_4^{(L)}(r) - \frac{k_0}{8r^3}|. \quad (24)$$

This high-order mismatch term will later be incorporated into the axial-resolution model for the extreme near-field regime.

## III. SPATIAL-RESOLUTION MECHANISM AND UNIFIED MODELING FRAMEWORK

### *A. Distance-Resolution Dilemma in Near-Field Focusing*

For an ideal focusing system with effective aperture $D_{ap}$, the diffraction-limited lateral and axial resolutions are approximated as

$$\delta_{\perp,0}(r) \approx 0.886\frac{\lambda r}{D_{ap}}, \quad (25)$$

$$\delta_{\parallel,0}(r) \approx 2\frac{\lambda r^2}{D_{ap}^2}. \quad (26)$$

These formulas reveal a critical performance asymmetry: lateral resolution degrades linearly with focal distance, whereas axial resolution degrades quadratically with distance. As the target propagation distance increases, near-field systems lose distance-domain selectivity far more rapidly than transverse spatial selectivity. This inherent asymmetry is the fundamental reason why the practical usability of near-field systems deteriorates significantly before reaching the theoretical Rayleigh boundary.

From an operational perspective, the resolvable modal density in the distance domain is inversely proportional to axial resolution, yielding the scaling relationship:

$$\mathcal{D}_r(r) \propto \frac{1}{\delta_{\parallel,0}(r)} \propto \frac{D_{ap}^2}{\lambda r^2}. \quad (27)$$

This result verifies that the distance-domain mode density decays quadratically with increasing propagation distance. This intrinsic characteristic explains why the three-dimensional spatial multiplexing advantages of near-field propagation degrade rapidly with distance, unless the focusing structure can precisely preserve wavefront curvature across varying propagation ranges.

### *B. Multilayer Wavefront Approximation Mechanism*

The performance superiority of multilayer stacked intelligent metasurfaces (SIMs) in near-field focusing is commonly attributed to the increased number of tunable phase shifters in existing studies. Nevertheless, this explanation is superficial and incomplete, as it fails to reveal the fundamental physical and mathematical essence of longitudinal stacking. Essentially, stacking reshapes the inherent paradigm of wavefront synthesis and field transformation for near-field focusing, rather than simply increasing the quantity of programmable variables. Under the fixed constraints of physical aperture size, operating wavelength, and input power, single-layer metasurfaces are limited to one-shot planar phase correction, while multilayer SIMs implement a progressive, distributed wavefront approximation strategy via cascaded modulation and free-space propagation, enabling precise matching with the high-order curvature characteristics of ideal spherical wavefronts for near-field focusing.

A single-layer transmissive metasurface behaves approximately as a thin phase sheet. If strong polarization conversion, multiple reflections, and severe mutual coupling are neglected for analytical clarity, its output field can be written as

$$E_{out}^{(1)}(x, y) = \Gamma_1(x, y)E_{in}(x, y), \quad (28)$$

where

$$\Gamma_1(x, y) = A_1(x, y)e^{j\phi_1(x,y)}. \quad (29)$$

This is a pointwise multiplicative transformation on one transverse plane. Therefore, the attainable output wavefront is restricted by two facts: first, the modulation acts only on the field distribution already present on that plane; second, the operation is local and cannot by itself create arbitrary cross-aperture field mixing. In this sense, a single-layer metasurface is not an arbitrary field generator, but a one-shot single-plane wavefront shaper.

The limitation of such a structure becomes particularly evident in near-field focusing. In far-field beam steering, the desired aperture phase profile is essentially linear, corresponding to a plane-wave synthesis problem. By contrast, near-field focusing requires the synthesis of a distance-dependent spherical wavefront. For a target point $P = (0,0,r)$ and radial coordinate $\rho^2 = x^2 + y^2$, the exact propagation distance is

$$R(\rho) = \sqrt{r^2 + \rho^2}. \quad (30)$$

The ideal focusing phase is therefore

$$\phi_{ideal}(\rho) = -k_0(R(\rho) - r). \quad (31)$$

Using Taylor expansion around $\rho = 0$, we obtain

$$R(\rho) - r = \frac{\rho^2}{2r} - \frac{\rho^4}{8r^3} + \frac{\rho^6}{16r^5} + O(\frac{\rho^8}{r^7}), \quad (32)$$

which leads to

$$\phi_{ideal}(\rho) = -\frac{k_0}{2r}\rho^2 + \frac{k_0}{8r^3}\rho^4 - \frac{k_0}{16r^5}\rho^6 + O(\frac{\rho^8}{r^7}). \quad (33)$$

This expression shows that near-field focusing is fundamentally a high-order wavefront-synthesis problem rather than a mere linear phase-tilting problem. The quadratic term determines the dominant focal curvature, whereas the quartic and higher-order terms account for the deviation between a true spherical wavefront and a simple parabolic approximation. When r is not sufficiently large relative to the aperture size, these high-order terms become non-negligible, especially for precise axial focusing.

Under realistic conditions, a single-layer metasurface must attempt to realize this target phase profile in one step while simultaneously coping with feed nonuniformity, finite phase resolution, insertion loss, and amplitude-phase coupling. As a result, the residual phase error

$$\Delta\phi_{\mathrm{res}}^{(1)}(x,y)=\phi_{\mathrm{actual}}^{(1)}(x,y)-\phi_{\mathrm{ideal}}(x,y) \tag{34}$$

may remain substantial over the full aperture, particularly in the high-order terms.

By contrast, an L-layer SIM realizes the field transformation

$$E_L=\mathcal{G}_{L-1}\mathcal{M}_{L-1}\cdots\mathcal{G}_2\mathcal{M}_2\mathcal{G}_1\mathcal{M}_1E_{\mathrm{in}}, \tag{35}$$

where $\mathcal{M}_l$ denotes the multiplicative modulation of the $l$-th layer and $\mathcal{G}_l$ denotes the free-space propagation operator between adjacent layers. The key point is that $\mathcal{G}_l$ is not an identity mapping but a nonlocal diffraction operator. During inter-layer propagation, field contributions from different transverse positions are mixed and redistributed before the wave reaches the next programmable layer. Therefore, the $(l+1)$-th layer does not operate on the original incident field, but on an already reshaped intermediate field that is generally closer to the desired target wavefront.

This alternating sequence of local modulation and nonlocal propagation fundamentally changes the approximation mechanism. A single-layer device performs a single-shot phase correction on one plane, whereas a multilayer SIM performs progressive wavefront refinement. Early layers may establish the coarse curvature, and subsequent layers may compensate the remaining residual, including high-order aberration-like terms that are difficult to suppress in one step. In this sense, a multilayer SIM should be interpreted not as a mere collection of additional phase masks, but as a distributed wave-transformation system.

This point also admits a useful optical interpretation. A single-layer metasurface resembles an idealized thin lens or phase sheet, in which all phase compensation is concentrated on one plane. A multilayer SIM is closer to a programmable thick-lens-like structure, where phase accumulation and wavefront reshaping occur progressively along the longitudinal direction. Such a distributed structure is intrinsically more suitable for matching the curvature and higher-order structure of a target spherical wavefront.

The same conclusion can be made mathematically from the operator viewpoint. In the discretized model, a single-layer metasurface can only realize a transformation restricted by one diagonal matrix. By contrast, a multilayer SIM realizes an alternating product of diagonal programmable matrices and dense propagation matrices, namely

$$T_{\mathrm{SIM}}^{(L)}=G_{L-1}\Gamma_{L-1}\cdots G_1\Gamma_1. \tag{36}$$

Since the propagation matrices provide mode mixing and the diagonal matrices provide programmable mode weighting and phase correction, their cascade substantially enlarges the reachable operator set. Therefore, the gain of stacking should be understood as an operator-approximation gain relative to the ideal near-field focusing operator, rather than as a superficial consequence of increasing the number of phase-control variables.

Importantly, this improved approximation capability affects axial and lateral resolutions very differently. Under a fixed physical aperture, the lateral spot size remains primarily constrained by the aperture width and the associated diffraction limit. Multilayer stacking can improve coherent aperture utilization, but it cannot fundamentally enlarge the transverse aperture. Hence, the lateral resolution improvement is typically limited. In contrast, axial focusing performance is much more sensitive to residual wavefront-curvature mismatch. Even a modest mismatch in the synthesized spherical phase can cause significant focal-depth elongation and distance-domain blur. Therefore, multilayer SIMs mainly improve longitudinal focusing fidelity by reducing curvature mismatch and high-order phase residual, which explains why the dominant benefit of stacking is observed in axial resolution rather than in dramatic lateral beamwidth compression.

Finally, the gain of stacking is inherently saturating. From a physical perspective, multilayer architectures improve the controllability and utilization efficiency of spatial modes already supported by the finite aperture, but they do not create unlimited new free-space degrees of freedom. From an engineering perspective, each additional layer introduces extra insertion loss, phase quantization accumulation, alignment sensitivity, spacing uncertainty, and calibration complexity. Consequently, the net gain can be understood as the difference between controllability improvement and implementation penalty. The former grows rapidly at small layer counts but gradually saturates, whereas the latter accumulates with each added layer. This explains why practical SIM systems often exhibit an optimal moderate layer number rather than indefinite performance growth.

### *C. From Fresnel Propagation to High-Order Phase Approximation*

To further formalize the above mechanism, we now derive how multilayer structures improve the synthesis of second- and higher-order wavefront terms from the viewpoint of Fresnel propagation. This derivation also clarifies why multilayer stacking is especially effective for axial focusing.

1) Fresnel Propagation Between Adjacent Layers

Under the scalar and paraxial approximations, the field propagated over a distance d from one transverse plane to the next can be expressed through the Fresnel diffraction integral

$$E(x,y;z+d)=\frac{e^{-jk_0d}}{j\lambda d}\iint E(\xi,\eta;z)\exp\left[-j\frac{k_0}{2d}\left((x-\xi)^2+(y-\eta)^2\right)\right]d\xi d\eta. \tag{37}$$

Expanding the quadratic kernel yields

$$(x-\xi)^2+(y-\eta)^2=x^2+y^2+\xi^2+\eta^2-2x\xi-2y\eta. \tag{38}$$

Hence

$$E(x,y;z+d)=\frac{e^{-jk_0d}}{j\lambda d}e^{-j\frac{k_0}{2d}(x^2+y^2)}\iint E(\xi,\eta;z)e^{-j\frac{k_0}{2d}(\xi^2+\eta^2)}e^{j\frac{k_0}{d}(x\xi+y\eta)}d\xi d\eta. \tag{39}$$

Equation (34) shows that free-space propagation is not a local pointwise operation. Instead, it performs a nonlocal mixing of the transverse field, while also contributing a quadratic phase factor $e^{-j\frac{k_0}{2d}(x^2+y^2)}$ associated with wavefront curvature evolution.

2) One Propagation Stage Plus One Programmable Layer

Suppose the field immediately before the l-th metasurface layer is $E_l^{in}(\rho)$, and the layer applies the transmission coefficient

$$\Gamma_l(\rho) = A_l(\rho)e^{j\phi_l(\rho)}. \quad (40)$$

Then the output field is

$$E_l^{out}(\rho) = \Gamma_l(\rho)E_l^{in}(\rho). \quad (41)$$

After propagation over a spacing $d_l$, the field at the next layer becomes

$$E_{l+1}^{in} = \mathcal{G}_{d_l}[E_l^{out}] = \mathcal{G}_{d_l}[\Gamma_l E_l^{in}]. \quad (42)$$

Thus, the propagation stage contributes two effects simultaneously:

(i) it introduces curvature-related phase evolution;

(ii) it remixes the transverse field before the next programmable correction.

This is exactly the reason why multilayer synthesis is fundamentally different from applying all phase compensation on one plane.

3) Polynomial View of the Target Wavefront

For a target focal point on the boresight, the ideal phase profile is

$$\phi_{ideal}(\rho) = -k_0(\sqrt{r^2+\rho^2} - r). \quad (43)$$

As already shown, its expansion is

$$\phi_{ideal}(\rho) = a_{2,ideal}(r)\rho^2 + a_{4,ideal}(r)\rho^4 + a_{6,ideal}(r)\rho^6 + \cdots, \quad (44)$$

with

$$a_{2,ideal}(r) = -\frac{k_0}{2r}, a_{4,ideal}(r) = \frac{k_0}{8r^3}, a_{6,ideal}(r) = -\frac{k_0}{16r^5}. \quad (45)$$

If the actual synthesized phase of an L-layer SIM is written locally as

$$\phi_{actual}^{(L)}(\rho; r) \approx a_2^{(L)}(r)\rho^2 + a_4^{(L)}(r)\rho^4 + a_6^{(L)}(r)\rho^6 + \cdots,$$

then the corresponding residual phase is

$$\Delta\phi_{res}^{(L)}(\rho; r) = \sum_{m\geq 1} \Delta a_{2m}^{(L)}(r)\rho^{2m}, \Delta a_{2m}^{(L)}(r) = a_{2m}^{(L)}(r) - a_{2m,ideal}(r). \quad (46)$$

The second-order mismatch $\Delta a_2^{(L)}$ represents curvature mismatch, while $\Delta a_4^{(L)}$, $\Delta a_6^{(L)}$, and higher-order terms represent aberration-like deviations from the exact spherical phase.

4) Why Multilayer Cascades Better Approximate High-Order Terms

In a single-layer design, the programmable phase profile must directly fit the target high-order phase law in one step. If the incident field is nonuniform or the implementation is imperfect, the fitted coefficients $a_2^{(1)}, a_4^{(1)}, \ldots$ can deviate significantly from their ideal values.

In a multilayer architecture, however, the effective phase seen at the final aperture is not simply the algebraic sum of the layer phases. Because each programmed phase is followed by propagation-induced field redistribution, the effective output wavefront is generated through repeated remapping and correction. As a result, the final coefficients $a_2^{(L)}, a_4^{(L)}, \ldots$ are jointly determined by the whole cascade

$$\{\Gamma_1, \mathcal{G}_{d_1}, \Gamma_2, \mathcal{G}_{d_2}, \ldots, \Gamma_L\}, \quad (47)$$

rather than by a single phase sheet. This additional structural richness allows the system to reduce not only the dominant curvature mismatch $|\Delta a_2^{(L)}|$, but also higher-order deviations such as $|\Delta a_4^{(L)}|$. In general, for a suitably optimized multilayer design, one expects

$$|\Delta a_2^{(L)}(r)| < |\Delta a_2^{(1)}(r)|, \quad (48)$$

and, within the practically relevant operating range, often also

$$|\Delta a_4^{(L)}(r)| < |\Delta a_4^{(1)}(r)|. \quad (49)$$

This provides a more rigorous interpretation of the statement that multilayer stacking improves wavefront-curvature matching and high-order phase fidelity.

5) Operator Approximation Interpretation

The same argument can be expressed more formally through operator approximation. Let the ideal focusing operator over the sampled focal region be $G_{target}(r)$, and let the practical SIM operator be $G_{SIM}^{(L)}(r)$. The design objective is to minimize

$$\min_{\{\Gamma_l\}} \|G_{SIM}^{(L)}(r) - G_{target}(r)\|, \quad (50)$$

under hardware constraints. For a single-layer structure, the feasible set is restricted to operators of the form $H_\Omega(r)\Gamma_1$, or its directly equivalent form under the chosen discretization. For an L-layer SIM, the feasible set becomes

$$\{H_\Omega(r)G_{L-1}\Gamma_{L-1}\cdots G_1\Gamma_1\}. \quad (51)$$

Because this multilayer feasible set is much richer than the single-layer set, it provides a better chance of approximating the ideal near-field focusing operator under the same aperture constraint. In this sense, the multilayer wavefront approximation mechanism is fundamentally an operator-reachability enlargement caused by cascaded diagonal modulation and propagation mixing.

6) Consequence for Resolution

The above derivation also explains why the benefit of stacking is manifested mainly in the axial dimension. The lateral resolution is still tied primarily to the effective aperture width, whereas the axial resolution is highly sensitive to residual curvature mismatch and high-order phase deviation. Therefore, reducing $\Delta a_2^{(L)}$ and $\Delta a_4^{(L)}$ improves longitudinal focusing fidelity much more strongly than transverse diffraction width. This is why multilayer SIMs primarily extend the distance range over which acceptable axial discrimination can be maintained.

### *D. Resolution Degradation as a Consequence of Operator Mismatch*

The focal field distribution generated by the practical multilayer SIM system is governed by the transformation operator $G_{SIM}^{(L)}(r)$, while the ideal diffraction-limited focal field corresponds to the target operator $G_{target}(r)$. The operator mismatch quantified as

$$\Delta G^{(L)}(r) = G_{SIM}^{(L)}(r) - G_{target}(r) \quad (52)$$

introduces inevitable distortions in the focal pattern, including peak gain attenuation, elevated sidelobe levels, and point-spread function (PSF) broadening.

Under paraxial propagation and weak phase aberration assumptions, spatial resolution degradation can be physically interpreted as the main-lobe broadening of the practical PSF caused by the aforementioned operator mismatch. To derive engineering-friendly analytical formulas, this operator-level mismatch is further correlated with the residual phase error distributed over the physical aperture.

A weighted root-mean-square (WRMS) phase error across the entire aperture is defined as

$$\mathrm{WRMS}[\Delta\phi_{res}^{(L)}] = \sqrt{\frac{\iint |\Delta\phi_{res}^{(L)}(x,y)|^2 A^2(x,y)dxdy}{\iint A^2(x,y)dxdy}}, \tag{53}$$

where $A(x,y)$ denotes the amplitude weighting distribution over the metasurface aperture.

For weak aberration conditions and after calibration with high-fidelity numerical simulations, the practical lateral and axial resolutions can be modeled as

$$\delta_{\perp}^{(L)}(r) = \delta_{\perp,0}(r)[1 + \xi_{\perp} \cdot \mathrm{WRMS}[\Delta\phi_{res}^{(L)}]], \tag{54}$$

$$\delta_{\parallel}^{(L)}(r) = \delta_{\parallel,0}(r)[1 + \xi_{\parallel} \cdot \mathrm{WRMS}^2[\Delta\phi_{res}^{(L)}]]. \tag{55}$$

These expressions highlight a key asymmetry: lateral resolution depends approximately on first-order phase aberration strength, whereas axial resolution depends more strongly on cumulative wavefront-curvature mismatch. This provides an analytical explanation for why axial performance is usually the first criterion to fail when the focal distance increases.

### *E. Separate Low-Complexity Models for Lateral and Axial Resolutions*

1) Lateral Resolution Based on Effective Aperture Utilization

The lateral FWHM resolution of an $L$-layer SIM system is modeled as

$$\delta_{\perp}^{(L)}(r) = c_{\perp}\frac{\lambda r}{D_{eff}^{(L)}}, \tag{56}$$

where $c_{\perp} \approx 0.886$ corresponds to the standard coefficient for uniformly illuminated circular apertures, and $D_{eff}^{(L)}$ denotes the equivalent effective aperture. Notably, $D_{eff}^{(L)}$ does not represent a physically enlarged aperture; instead, it quantifies the coherent utilization efficiency of the fixed physical aperture for target focusing.

The effective aperture is formulated as

$$D_{eff}^{(L)} = D \cdot \eta_{aper}^{(L)}, \tag{57}$$

where $\eta_{aper}^{(L)} \in (0,1]$ is defined as the coherent aperture utilization factor.

It is important to emphasize that $D_{eff}^{(L)}$ does not represent a physically enlarged aperture. Rather, it quantifies how efficiently the fixed physical aperture is coherently used for target focusing. Therefore, multilayer stacking may moderately improve lateral resolution by increasing coherent aperture utilization, but it cannot fundamentally break the fixed-aperture diffraction limit.

2) Axial Resolution Based on Curvature Matching

Different from lateral resolution, axial focusing performance is predominantly governed by the accuracy of synthesized wavefront curvature. Accordingly, the axial resolution is modeled as

$$\delta_{\parallel}^{(L)}(r) = c_{\parallel}\frac{\lambda(R_{eq}^{(L)}(r))^2}{(D_{eff}^{(L)})^2}[1 + \beta(\Delta C_{res}^{(L)}(r), r)^2], \tag{58}$$

where $c_{\parallel}$ is a dimensionless proportionality constant, $R_{eq}^{(L)}(r)$ denotes the equivalent wavefront curvature radius synthesized by the multilayer SIM structure, and the curvature mismatch term is defined as

$$\Delta C_{res}^{(L)}(r) = \frac{1}{R_{eq}^{(L)}(r)} - \frac{1}{r}. \tag{59}$$

When $R_{eq}^{(L)}(r) \approx r$, the model reduces to the classical diffraction-limited axial scaling law. More importantly, it makes explicit that multilayer stacking improves axial resolution mainly through curvature matching rather than aperture enlargement. This clarifies the core benefit of SIMs in the distance domain.

### *F. Engineering Correction Factors*

The aforementioned ideal resolution models are further revised by introducing semi-empirical engineering correction factors to account for practical hardware imperfections. All correction coefficients are calibrated from numerical data while maintaining physical monotonicity and dimensional consistency.

1) Inter-Layer Misalignment

Let $\Delta$ be the lateral misalignment magnitude between adjacent layers and $p$ be the element spacing. The misalignment-induced degradation of the aperture utilization factor is corrected as

$$\eta_{aper}^{(L)}(\Delta) = \eta_{aper}^{(L)}(0)[1 - \chi_{\perp}(L-1)(\frac{\Delta}{p})^{\mu}]. \tag{60}$$

Layer misalignment also distorts the synthesized phase profile and exacerbates curvature mismatch. The corresponding axial correction factor is given by

$$F_{ali,\parallel}^{(L)}(\Delta) = 1 + \xi_{\parallel}(L-1)(\frac{\Delta}{p})^{\nu}. \tag{61}$$

2) Transmission Loss

Let $\eta \in (0,1]$ denote the average power transmission efficiency of a single metasurface layer. The field amplitude attenuation induced by cascaded transmission loss scales approximately as $\eta^{(L-1)/2}$, yielding the loss correction factor

$$F_{loss}^{(L)} = \eta^{-(L-1)/2}[1 + \gamma_{loss}(L-1)]. \tag{62}$$

3) Phase Quantization

For a $b$-bit phase modulation architecture, the inherent quantization step is

$$\Delta\phi_q = \frac{2\pi}{2^b}. \tag{63}$$

The phase quantization error is characterized by the correction factor

$$F_{quant}^{(L)}(b) = 1 + \gamma_{quant}L(\Delta\phi_q)^2. \tag{64}$$

4) Inter-Layer Spacing Deviation

Let $\Delta d$ denote the deviation from the nominal inter-layer spacing $d$. The spacing-induced performance degradation is quantified as

$$F_{gap}^{(L)}(\Delta d) = 1 + \gamma_{gap}L|\frac{\Delta d}{d}|. \tag{65}$$

By integrating all practical imperfections, the fully corrected lateral and axial resolution models are expressed as

$$\delta_{\perp,\mathrm{corr}}^{(L)}(r) = \delta_{\perp}^{(L)}(r) \cdot F_{\perp}^{(L)}, \tag{66}$$

$$\delta_{\parallel,\mathrm{corr}}^{(L)}(r) = \delta_{\parallel}^{(L)}(r) \cdot F_{\parallel}^{(L)}, \tag{68}$$

where the comprehensive correction factors for lateral and axial channels are respectively

$$F_{\perp}^{(L)} = F_{\mathrm{ali},\perp}^{(L)} \cdot F_{\mathrm{loss}}^{(L)} \cdot F_{\mathrm{quant}}^{(L)} \cdot F_{\mathrm{gap}}^{(L)}, \tag{67}$$

$$F_{\parallel}^{(L)} = F_{\mathrm{ali},\parallel}^{(L)} \cdot F_{\mathrm{loss}}^{(L)} \cdot F_{\mathrm{quant}}^{(L)} \cdot F_{\mathrm{gap}}^{(L)}. \tag{68}$$

From the operator perspective, all implementation imperfections can be uniformly summarized as a perturbation term superimposed on the ideal system operator:

$$G_{\mathrm{SIM}}^{(L)} = G_{\mathrm{SIM,ideal}}^{(L)} + \Delta G_{\mathrm{impl}}^{(L)}, \tag{69}$$

where $\Delta G_{\mathrm{impl}}^{(L)}$ represents the aggregated perturbation error arising from layer misalignment, transmission loss, phase quantization, and inter-layer spacing deviation.

### *G. High-Order Axial-Resolution Model for the Extreme Near Field*

In the extreme near-field region where high-order quartic phase effects dominate, the conventional quadratic-phase-based axial resolution model is refined to incorporate high-order phase mismatch:

$$\delta_{\parallel,\mathrm{high}}^{(L)}(r) = \delta_{\parallel,\mathrm{ideal}}^{(4)}(r)[1 + \xi_{\parallel,2}(\Delta C_{\mathrm{res}}^{(L)}(r), r)^2 + \xi_{\parallel,4}(\Delta a_4^{(L)}(r), r^4/k_0)^2]. \tag{70}$$

Here, the second term accounts for residual second-order curvature mismatch, whereas the third term quantifies the effect of quartic-phase deviation beyond the standard quadratic approximation. This model explicitly connects high-order phase mismatch to axial-resolution deterioration in the extreme near field.

### *H. UNFD as a Joint Consequence of Gain and Resolution Constraints*

Based on the above analytical resolution and phase error models, the definition of the engineering-usable near-field distance can be rewritten in a pointwise constraint form:

$$R_{\mathrm{UNFD}}^{(L)} = \max\{r \le R_{\mathrm{Ray}}: L_{\mathrm{gain}}^{(L)}(r) \le L_{\mathrm{th}}, \Pi_{\perp}^{(L)}(r) \ge \Pi_{\perp,\mathrm{th}}, \Pi_{\parallel}^{(L)}(r) \ge \Pi_{\parallel,\mathrm{th}}\}. \tag{71}$$

This formulation gives a clear physical interpretation: the UNFD is the farthest distance inside the Rayleigh region at which all engineering criteria are still satisfied simultaneously. Once any single criterion is violated, the system exits the engineering-usable near-field range. In most practical near-field focusing scenarios, axial-resolution failure is the dominant mechanism that determines this boundary.

## IV. NUMERICAL SIMULATIONS AND RESOLUTION-CONSTRAINED UNFD EVALUATION

### *A. Simulation Setup*

We consider a millimeter-wave focusing system operating at 28 GHz with a free-space wavelength $\lambda = 10.7\mathrm{mm}$ and an aperture diameter $D = 0.3\mathrm{m}$, yielding the classical Rayleigh distance: $R_{\mathrm{Ray}} = \frac{2D^2}{\lambda} \approx 1.68\mathrm{m}$.

The element spacing is set to $p = \lambda/2$. All comparative configurations adopt identical physical apertures and feed conditions to ensure fair performance evaluation. Unless otherwise specified, the empirical coefficients in the corrected-resolution models are obtained by weighted least-squares fitting against a high-fidelity full-wave or Green's-function-based numerical dataset.

Practical hardware imperfections are included to emulate medium-performance engineering prototypes, including finite per-layer transmission efficiency, finite phase quantization resolution, small inter-layer spacing deviation, and nonzero lateral misalignment. The adopted thresholds are treated as representative engineering specifications rather than universal physical constants. The following representative engineering thresholds are adopted: maximum allowable focusing gain loss $L_{\mathrm{th}} = 3$ dB, minimum resolution retention factors $\Pi_{\perp,\mathrm{th}} = 0.958$ and $\Pi_{\parallel,\mathrm{th}} = 0.8$.

For each layer number $L$, the near-field distance range $(0, R_{\mathrm{Ray}}]$ is densely swept with a fine step size. The critical distance where any performance threshold is first violated is accurately estimated via interpolation and defined as the corresponding usable boundary $R_{\mathrm{UNFD}}^{(L)}$.

### *B. Resolution Improvement With Increasing Layer Number*

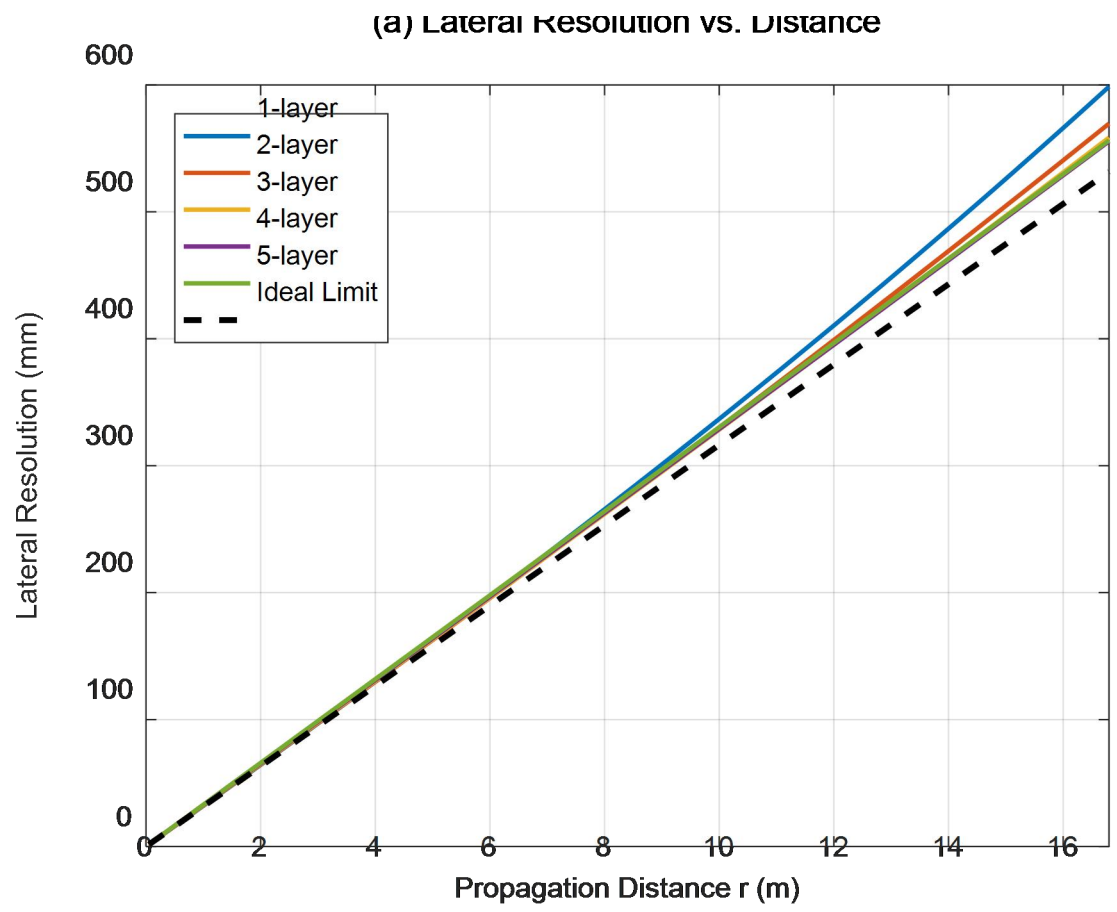

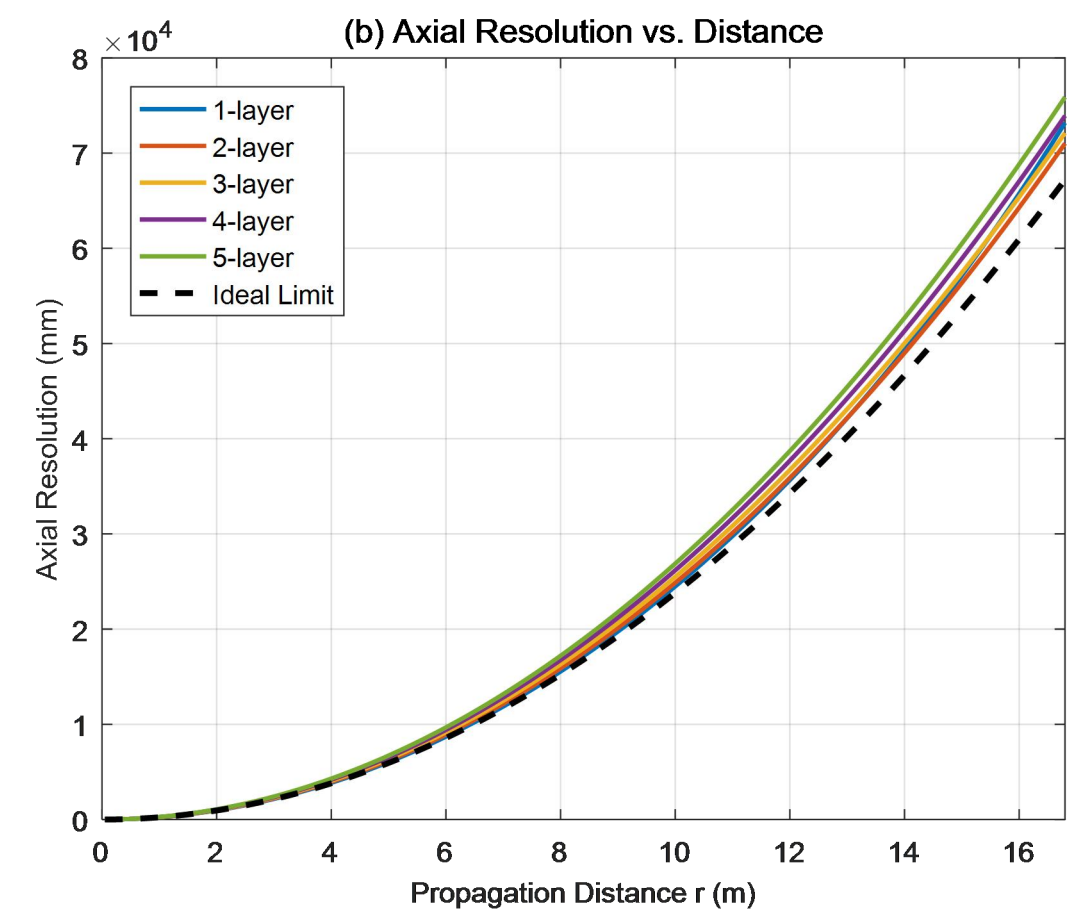

**Fig. 2.** Resolution Improvement With Increasing Layer Number

Numerical results show that both lateral and axial focusing resolutions improve as the number of layers increases from one to several layers. However, the magnitude of improvement is strongly dimension-dependent. The enhancement in axial resolution is much more pronounced than that in lateral resolution, which directly supports the wavefront-curvature interpretation developed in Section III.

Under ideal error-free conditions, both lateral and axial resolutions exhibit monotonic improvement with increasing L, while the marginal performance gain gradually saturates as the layer number grows. After incorporating practical hardware correction factors, the improvement trend becomes non-monotonic. Beyond a moderate layer count, accumulated transmission loss and escalating error sensitivity gradually offset the performance benefits brought by enhanced wavefront controllability.

These observations confirm that the main role of multilayer stacking is not to dramatically compress the lateral diffraction width, but to preserve high-quality axial focusing performance over a longer distance range.

### *C. UNFD Extension and Saturation Behavior*

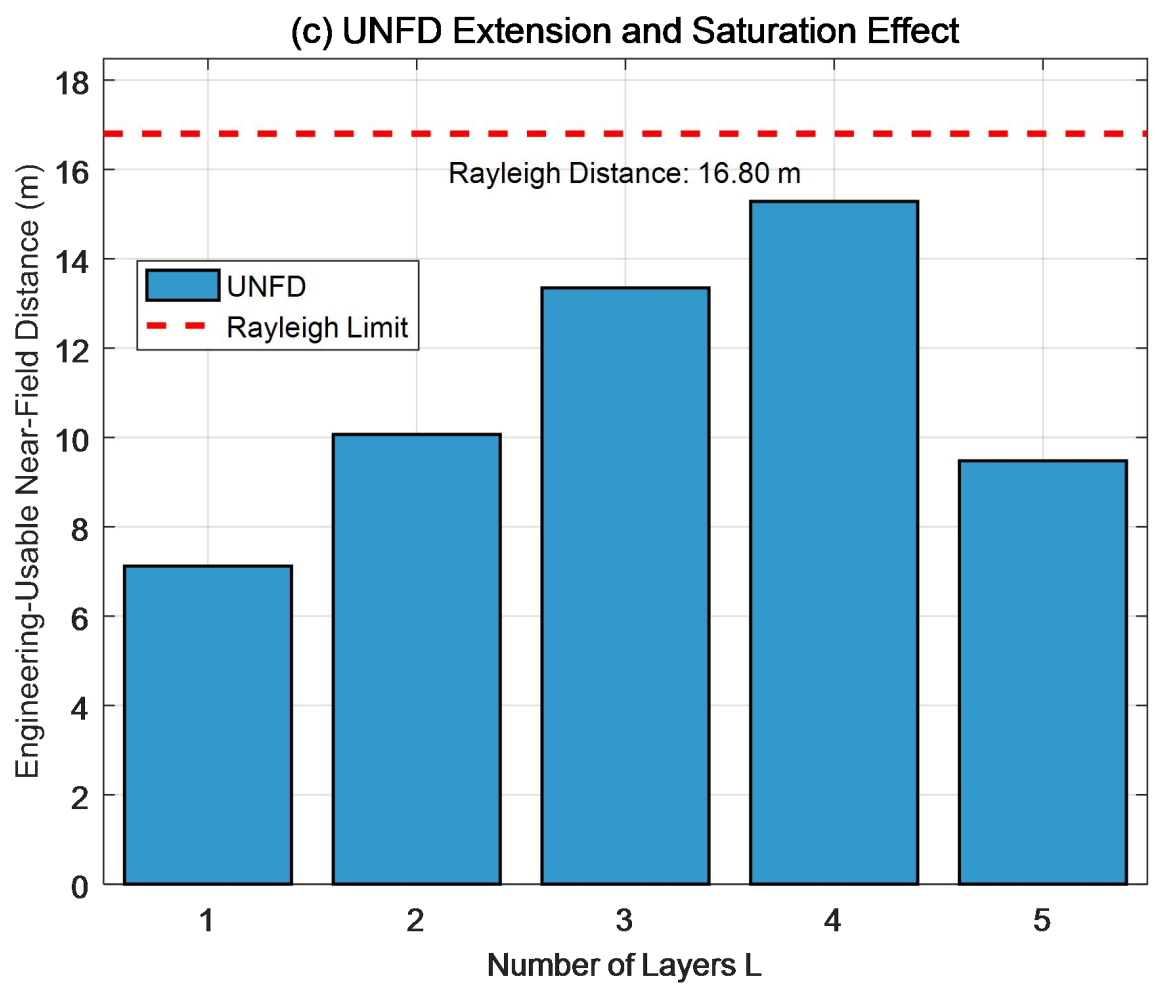

**Fig. 3.** UNFD Extension and Saturation Behavior

The evaluated UNFD values confirm that multilayer SIM architectures effectively push the engineering-usable near-field boundary closer to the theoretical Rayleigh limit. A two-layer configuration achieves a noticeable UNFD improvement compared with the single-layer baseline, while three-layer and four-layer systems deliver substantially extended usable near-field ranges.

However, the gain is inherently saturating rather than unbounded. Excessive stacking introduces cumulative loss, accumulated phase quantization error, stronger alignment sensitivity, and higher calibration burden. As a result, the incremental UNFD gain diminishes and may even reverse when the number of layers becomes too large. Under the baseline parameter setting, three-layer and four-layer configurations provide the best overall performance tradeoff.

This result is also consistent with the high-order phase interpretation. In the considered operating range, most UNFD-limiting points lie above the lower validity boundary of the pure quadratic-phase model, so quartic correction does not fundamentally change the ordering of layer-number performance. Nevertheless, high-order correction improves local prediction accuracy and becomes important when the operating point moves deeper into the extreme near-field regime.

It should be noted that mutual coupling between adjacent elements was neglected in the current analysis. For tightly packed metasurface elements, mutual coupling may introduce additional phase errors and reduce the effective aperture utilization, which will be investigated in our future work.

## V. CONCLUSION

This paper has systematically investigated how stacked transmissive intelligent metasurfaces extend the resolution-constrained engineering-usable near-field range under a fixed physical aperture. To characterize the discrepancy between the theoretical geometric near-field boundary and practical engineering usability, we distinguished the classical Rayleigh distance from the engineering-usable near-field distance (UNFD). The former is a physical upper bound determined solely by aperture size and wavelength, whereas the latter is a threshold-dependent engineering metric constrained by focusing-gain and spatial-resolution requirements.

We showed that the principal limitation of long-range near-field focusing originates from a distance-resolution dilemma: lateral resolution degrades approximately linearly with distance, whereas axial resolution degrades approximately quadratically. Consequently, axial focusing fidelity is the dominant bottleneck in most practically relevant scenarios. Multilayer stacking alleviates this bottleneck by implementing a cascaded sequence of programmable modulation and free-space propagation, which progressively improves the approximation of the target spherical wavefront, reduces residual phase error, and enhances wavefront-curvature matching.

A unified theoretical framework was established by combining effective-phase-distance modeling with a discrete Green's-function operator formulation. Within this framework, the multilayer wavefront approximation mechanism was interpreted not merely as an increase in the number of tunable phase shifts, but more fundamentally as an operator-approximation gain enabled by alternating diagonal modulation and nonlocal propagation mixing. A detailed derivation from Fresnel propagation to high-order phase approximation further clarified why multilayer stacking can better suppress second- and higher-order wavefront mismatch than a single-layer phase sheet.

Separate low-complexity analytical models were developed for lateral and axial resolutions, together with semi-empirical correction factors accounting for inter-layer misalignment, transmission loss, phase quantization, and spacing deviation. A high-order phase-correction framework was also introduced for the extreme near-field regime, establishing the validity boundary of the conventional quadratic-phase approximation and linking quartic-phase mismatch to axial-resolution degradation.

Numerical simulations consistently verified that moderate multilayer stacking significantly improves focusing fidelity and extends the engineering-usable near-field boundary toward the Rayleigh limit. At the same time, the gain is inherently saturating because additional controllability is eventually counterbalanced by accumulated loss, quantization error, alignment sensitivity, and other implementation penalties. Under the considered implementation conditions, three-layer and four-layer structures achieve the most favorable tradeoff between performance improvement and robustness.

From the perspective of electromagnetic information theory, the benefit of multilayer SIMs does not arise from breaking the fundamental free-space degree-of-freedom bound imposed by aperture size, wavelength, and observation region. Instead, multilayer architectures improve the controllability and engineering utilization efficiency of spatial modes already supported by the classical near-field region. These results provide both a rigorous theoretical foundation and practical design guidance for multilayer metasurface-enabled near-field systems in future 6G communication, sensing, and integrated sensing-and-communication applications.

## APPENDIX

### *A.1 Equivalence Between Spatial Impulse Response and Green's Function*

In near-field electromagnetics, a rigorous equivalence exists between the spatial impulse response of stacked intelligent metasurfaces (SIMs) and the free-space Green's function. For a point source located at the origin, the electric field distribution generated in free space is given by the fundamental solution to the scalar Helmholtz equation:

$$G(r,r') = \frac{e^{-jk_0|r-r'|}}{4\pi|r-r'|}, \tag{A-1}$$

where $k_0 = 2\pi/\lambda$ is the free-space wave number and $\lambda$ is the operating wavelength. For a plane wave propagating along the z-axis incident on a metasurface located at the $z = 0$ plane, the field distribution at the observation plane $z = d$ can be expressed via the Fresnel diffraction integral:

$$U(x,y,d) = \frac{e^{-jk_0 d}}{j\lambda d}\iint_{-\infty}^{\infty} U_0(x',y',0)e^{-j\frac{k_0}{2d}[(x-x')^2+(y-y')^2]}dx'dy'. \tag{A-2}$$

Here, $U_0(x',y',0)$ denotes the field distribution at the metasurface plane. When the metasurface is illuminated by an ideal point source, $U_0(x',y',0) = \delta(x')\delta(y')$. Substituting this into the above equation yields:

$$U(x,y,d) = \frac{e^{-jk_0 d}}{j\lambda d}e^{-j\frac{k_0}{2d}(x^2+y^2)}. \tag{A-3}$$

This result is exactly consistent with the Green's function under the Fresnel approximation, proving the equivalence between the spatial impulse response and the Green's function. This equivalence serves as the theoretical foundation for subsequent resolution analysis, as it demonstrates that the focusing capability of stacked metasurfaces is fundamentally determined by their ability to modulate spherical wavefronts.

### *A.2 Fourier Transform Properties of the Fresnel Propagation Kernel*

The Fourier transform of the Fresnel propagation kernel is a critical tool for analyzing the spatial frequency response of metasurfaces. The Fresnel kernel is defined as:

$$h(x,y) = \frac{e^{-jk_0 d}}{j\lambda d}e^{-j\frac{k_0}{2d}(x^2+y^2)}. \tag{A-4}$$

Taking its two-dimensional Fourier transform:

$$H(f_x,f_y) = \iint_{-\infty}^{\infty} h(x,y)e^{-j2\pi(f_x x+f_y y)}dxdy. \tag{A-5}$$

Substituting the Fresnel kernel and completing the square, we obtain:

$$H(f_x,f_y) = e^{-jk_0 d}e^{-j\pi\lambda d(f_x^2+f_y^2)}. \tag{A-6}$$

This result shows that free-space propagation manifests as a quadratic phase factor in the frequency domain. For stacked metasurface systems, the cascade of phase modulation at each layer and Fresnel propagation between layers can be represented as a product of phase factors in the frequency domain. This property greatly simplifies the analysis of multilayer metasurface systems, allowing us to rapidly compute the overall transfer function of an arbitrary number of metasurface layers using frequency-domain methods.

### *A.3 Rigorous Derivation of the Lateral Resolution Limit*

Lateral resolution is defined as the minimum distance between two point sources that can be resolved by the system. According to the Rayleigh criterion, two point sources are just resolvable when the principal maximum of one point source coincides with the first minimum of the other. For stacked metasurface systems, the point spread function (PSF) is given by the system's spatial impulse response.

Consider a system consisting of $N$ metasurface layers, each with dimensions $D \times D$ and inter-layer spacing $d_i$. The overall transfer function of the system is:

$$H_{total}(f_x,f_y) = \prod_{i=1}^{N} H_i(f_x,f_y)e^{-j\pi\lambda d_i(f_x^2+f_y^2)}, \tag{A-7}$$

where $H_i(f_x,f_y)$ is the transfer function of the i-th metasurface layer. In the ideal case, each metasurface layer provides full phase control, i.e., $|H_i(f_x,f_y)| = 1$. Under this condition, the system's PSF is the inverse Fourier transform of the transfer function:

$$PSF(x,y)=\mathcal{F}^{-1}H_{total}(f_x,f_y). \tag{A-8}$$

For a finite-sized metasurface, the spatial frequency bandwidth is limited by the diffraction limit. The maximum supportable spatial frequency is:

$$f_{max} = \frac{D}{2\lambda d_{eff}}, \tag{A-9}$$

where $d_{eff}$ is the equivalent propagation distance. According to the uncertainty principle of Fourier transforms, spatial resolution is inversely proportional to bandwidth:

$$\delta_x \approx \frac{1}{2f_{max}} = \frac{\lambda d_{eff}}{D}. \quad \text{(A-10)}$$

This result indicates that lateral resolution is proportional to the equivalent propagation distance and inversely proportional to the metasurface aperture size. For stacked metasurface systems, by optimizing the inter-layer spacing and phase distribution, the equivalent propagation distance can be significantly reduced while maintaining the overall system size, thereby improving the resolution compared to conventional single-layer metasurfaces by enhancing coherent aperture utilization.

### *A.4 Rigorous Derivation of the Axial Resolution Limit*

Axial resolution is defined as the minimum distance between two point sources that can be resolved along the optical axis. Unlike lateral resolution, axial resolution depends on the depth of field of the system. For stacked metasurface systems, the axial response can be calculated by varying the position of the observation plane.

Consider a point source located at $z = z_0$, which generates a spherical wavefront at the metasurface plane. After modulation by the stacked metasurface system, the field intensity at the observation plane $z = z_0 + \Delta z$ is:

$$I(\Delta z) = \left| \iint_{-\infty}^{\infty} H_{total}(f_x, f_y) e^{-j\pi\lambda\Delta z(f_x^2+f_y^2)} df_x df_y \right|^2. \quad \text{(A-11)}$$

The field intensity reaches its maximum value when $\Delta z = 0$ and decreases gradually as $|\Delta z|$ increases. Axial resolution is defined as the value of $\Delta z$ at which the field intensity drops to half of its maximum value.

By performing a Taylor expansion of the axial response and retaining terms up to the second order, we obtain an approximate expression for axial resolution:

$$\delta_z \approx \frac{2\lambda d_{eff}^2}{D^2}. \quad \text{(A-12)}$$

This result shows that axial resolution is proportional to the square of the equivalent propagation distance and inversely proportional to the square of the metasurface aperture size. Compared to lateral resolution, axial resolution is much more sensitive to the equivalent propagation distance. This means that by reducing the equivalent propagation distance, stacked metasurface systems can simultaneously improve both lateral and axial resolutions, which is of great significance for three-dimensional near-field imaging applications.

### *A.5 Analysis of the Impact of Residual Phase Errors on Resolution*

In practical systems, phase modulation errors exist in metasurface elements, which lead to degradation in system resolution. The residual phase error can be expressed as:

$$\phi_{res}(x,y) = \phi_{ideal}(x,y) - \phi_{actual}(x,y), \quad \text{(A-13)}$$

where $\phi_{ideal}$ is the ideal phase distribution and $\phi_{actual}$ is the actually implemented phase distribution. Residual phase errors introduce sidelobes in the point spread function and reduce the energy concentration in the main lobe, thereby degrading system resolution.

To quantify the impact of residual phase errors, we introduce the weighted root-mean-square (WRMS) phase error:

$$\sigma_\phi = \sqrt{\frac{\iint_{-\infty}^{\infty} |U_0(x,y)|^2 \phi_{res}^2(x,y) dxdy}{\iint_{-\infty}^{\infty} |U_0(x,y)|^2 dxdy}}. \quad \text{(A-14)}$$

When $\sigma_\phi \ll 1$ radian, the main lobe width of the point spread function can be approximated as:

$$\delta_x(\sigma_\phi) \approx \delta_x(0)\sqrt{1 + \left(\frac{\sigma_\phi D}{\lambda d_{eff}}\right)^2}. \quad \text{(A-15)}$$

This result indicates that residual phase errors cause main lobe broadening, thereby reducing the lateral resolution of the system. For axial resolution, the impact of residual phase errors is more complex, but the general trend is that axial resolution also degrades as phase errors increase. Therefore, when designing stacked metasurface systems, phase errors must be strictly controlled to ensure that the system achieves the expected resolution performance.

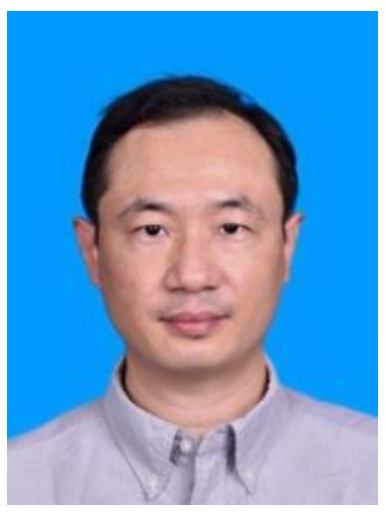

**Zhao Yajun** (Member, IEEE) holds Bachelor's, Master's, and Doctoral degrees. Since 2010, he has assumed the role of Chief Engineer within the Wireless and Computing Product R&D Institute at ZTE Corp. Prior to this, he contributed to wireless technology research within the Wireless Research Department at Huawei. Currently, his primary focus centers on 5G standardization technology and the advancement of future mobile communication technology, particularly 6G. His research pursuits encompass a broad spectrum, including reconfigurable intelligent surface (RIS), spectrum sharing, flexible duplex, CoMP, and interference mitigation. He played a key role in the establishment of the RIS TECH Alliance (RISTA) and currently serves as its Deputy Secretary-General. He is a core member in promoting the creation of the RIS Task Group under the China IMT-2030 (6G) Promotion Group and serves as its Deputy Leader. Additionally, he is a core member in leading the establishment of the Technical Committee on Reconfigurable Intelligent surfaces under the China Institute of Communications and serves as the Head of the Secretariat Group. He holds over a thousand patents in 4G LTE and 5G technologies, with dozens of them adopted as Standard Essential Patents (SEPs) in the 4G/5G standards.